\documentclass[letter,twocolumn]{jpsj2} 
\usepackage{lscape}

\title{On the Symmetry of Low-Field Ordered Phase of PrFe$_{4}$P$_{12~}$:$^{~31}$P NMR}

\author{Jun \textsc{Kikuchi}$^{1,2}$\thanks{E-mail address: jkiku@isc.meiji.ac.jp},
Masashi \textsc{Takigawa}$^{1}$, 
Hitoshi \textsc{Sugawara}$^{3,4}$
and Hideyuki \textsc{Sato}$^{4}$}

\inst{$^{1}$Institute for Solid State Physics, University of Tokyo, Kashiwa, Chiba 277-8581 \\
$^{2}$Department of Physics, Meiji University, Kawasaki, Kanagawa 214-8571 \\
$^{3}$Faculty of Integrated Arts and Science, University of Tokushima, Tokushima 770-8502 \\
$^{4}$Graduate School of Science, Tokyo Metropolitan University, Hachioji, Tokyo 192-0397}

\abst{We have performed $^{31}$P nuclear magnetic resonance (NMR) experiments on the 
filled skutterudite compound PrFe$_{4}$P$_{12}$ to investigate the exotic ordered phase 
at low temperatures and low magnetic fields. Analysis of the NMR line splitting in the ordered 
phase indicates that totally symmetric ($\Gamma_1$-type) staggered magnetic multipoles are 
induced by magnetic fields.  This is incompatible with any type of quadrupole order at zero field.  
We conclude that the ordered phase has broken translational symmetry with the wave 
vector $\mathbf{Q}=(1,0,0)$ but the $T_h$ point symmetry at the Pr sites is not broken 
by the non-magnetic $\Gamma_1$-type order parameter.}

\kword{PrFe$_{4}$P$_{12}$, NMR, skutterudite, multipole moment, order parameter, hyperfine interaction}

\begin{document}
\maketitle

Intermetallic compounds with the filled-skutterudite structure $RT_{4}X_{12}$ 
($R$ = rare earth, $T$ = transition metal, $X$ = pnictogen) has attracted strong recent attention 
because of a variety of intriguing phenomena in a common crystal structure such as 
metal-insulator transition~\cite{sekine97}, multipole ordering~\cite{yoshizawa041}, 
exotic superconductivity~\cite{bauer02}, and anomalous phonons~\cite{keppens98}.
Among them, PrFe$_{4}$P$_{12}$ shows a peculiar phase transition at 
$T_\mathrm{A}$=6.5~K~\cite{aoki02}.  The low temperature phase has no spontaneous magnetic 
moment at zero field~\cite{keller01} and is suppressed by a magnetic field of 4--7~T dependent 
on the field directions~\cite{aoki02}, resulting in a heavy-fermion state with a large cyclotron effective 
mass $m_c^*=81\,m_0$~\cite{sugawara02}.  The low temperature phase has a structural 
modulation with the wave vector $\mathbf{Q}=(1,0,0)$, indicating loss of $(\tfrac{1}{2},\tfrac{1}{2},\tfrac{1}{2})$
translation~\cite{iwasa02}.  Spatial ordering of distinct electronic states of two Pr$^{3+}$ ions 
in the bcc unit cell was also observed by resonant X-ray scattering,\cite{ishii03} and the field-induced 
staggered magnetization was observed by neutron scattering~\cite{hao03}. Although these experiments and the 
elastic measurements~\cite{nakanishi01} suggest an 
antiferro-quadrupole order likely to be of $\Gamma_{23}$-type, direct identification of the 
order parameter has not been made yet.  
 
In this letter, we report results of nuclear magnetic resonance (NMR) experiments on 
$^{31}$P nuclei (spin 1/2) in a single crystal of PrFe$_{4}$P$_{12}$,  focusing on the 
ordered phase. We observed field-induced splitting of the $^{31}$P NMR lines below $T_\mathrm{A}$. 
The results for various field directions revealed that the splitting is due to totally-symmetric 
staggered magnetic multipoles (octupole as well as dipole) belonging to the $\Gamma_1$ representation.   
We conclude that the order parameter at zero field has $\Gamma_1$ symmetry, 
excluding any type of quadrupole order.
\begin{figure}[b]
    \begin{center}
    \includegraphics[width=60mm]{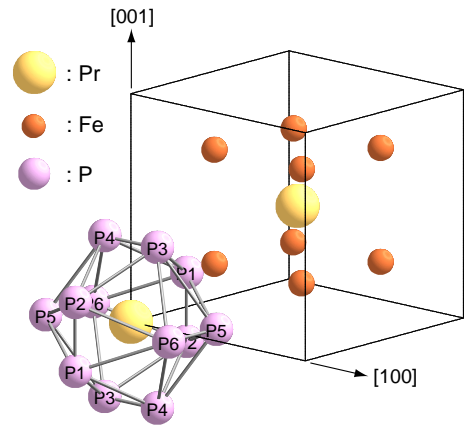}
    \caption{(Color online) Schematic view of the crystal structure of PrFe$_{4}$P$_{12}$.
    The cage surrounding the body-centered Pr atom is omitted for clarity.
    \label{fig:structure}}
    \end{center}
\end{figure}

In the filled skutterudite structure (the space group $Im\bar{3}$, $T_h^5$), Pr atoms are surrounded 
by an icosahedral cage of P atoms and form a body-centered cubic lattice as shown in 
Fig.~\ref{fig:structure}~\cite{keller01,jeitschko77}.  Iron atoms sit at the middle between the corner 
and the body-centered Pr atoms.  Although all the P sites are crystallographically equivalent, they have 
different NMR frequencies in magnetic field because of anisotropic hyperfine interaction.  
For later discussion, we define six types of P sites giving distinct NMR 
frequencies for general field directions as P1$(0,u,v)$, P2$(0,\bar{u},v)$, P3$(v,0,u)$, P4$(v,0,\bar{u})$, 
P5$(u,v,0)$ and P6$(\bar{u},v,0)$, where $u$ and $v$ are the asymmetry parameters (Fig.~\ref{fig:structure}).\cite{fn:site} 

\begin{figure}[t]
    \begin{center}
    \includegraphics[width=85mm]{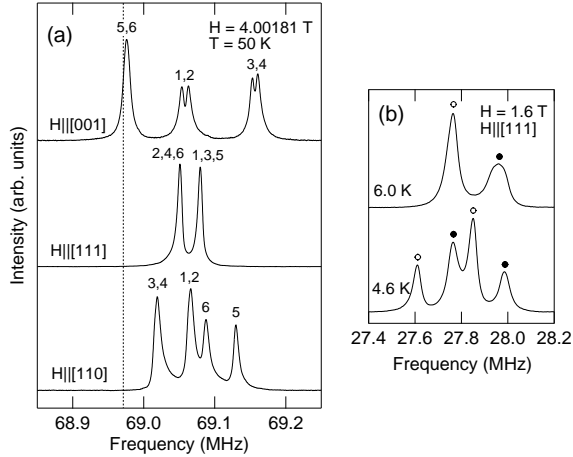}
    \caption{(a) $^{31}$P NMR spectra at 50~K obtained by Fourier transforming the spin-echo signals. 
      The numbers indicate the P sites (P1--P6) to which each peak is assigned. 
      The dotted line shows the reference frequency at zero shift. 
      (b) $^{31}$P NMR spectra at 6.0~K (above $T_\mathrm{A}$) and at 4.6~K (below $T_\mathrm{A}$) 
      for the field of 1.6~T along [111].  NMR lines from P1, P3 and P5 sites 
      (P2, P4 and P6 sites) are indicated by solid (open) circles. 
    \label{fig:spectra}}
    \end{center}
\end{figure}

If the direction of the external field  $\mathbf{H}$ is invariant under a symmetry operation which 
transforms one P site to another, those two P sites must have the same NMR frequency.  For example, 
P1, P3 and P5 sites (P2, P4, and P6 sites) transform each other by 120$^{\circ}$ rotation around 
[111], therefore, they should give a single NMR line for $\mathbf{H} \parallel [111]$. Likewise, 
for $\mathbf{H} \parallel [001]$, there should be three lines generated by three 
pairs of sites, (P1, P2), (P3, P4), and (P5, P6).  For $\mathbf{H} \parallel [110]$, P5 and P6 become
inequivalent and we expect four lines. Figure~\ref{fig:spectra}(a) shows the NMR spectra 
at $T$=50~K above $T_\mathrm{A}$.  
We observed three, four, and two lines for $\mathbf{H}$ parallel to the [001], [110], 
and [111] directions, respectively.  This agrees with the consequence of crystal symmetry.  
Detailed angle dependence of the resonance frequencies (not shown) 
leads to the site assignment shown in Fig. ~\ref{fig:spectra}(a).  The small splitting of the two 
high frequency lines for $\mathbf{H} \parallel [001]$ is probably due to nuclear spin-spin coupling.    
\begin{figure}[t]
    \begin{center}
    \includegraphics[width=70mm]{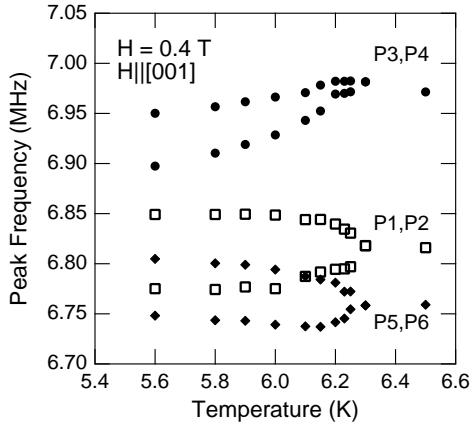}
    \caption{Temperature dependence of the NMR frequencies near $T_\mathrm{A}$=6.27~K for the field of 
    0.4~T along [001].
    \label{fig:splitvsT}}
    \end{center}
\end{figure}

On crossing $T_\mathrm{A}$ from above, each line splits into a pair of lines as shown in 
Figs.~\ref{fig:spectra}(b) and \ref{fig:splitvsT}.   The splitting grows continuously near 
$T_\mathrm{A}$ at low fields (0.4~T) as shown in Fig.~\ref{fig:splitvsT}.  At higher fields 
above about 1.5 T, however, the splitting develops discontinuously and there is a narrow 
temperature range in the vicinity of $T_\mathrm{A}$ where the spectrum has both split 
and unsplit lines.  This indicates that the second order phase transition 
at low fields changes to first order at higher fields, consistent with the results of specific-heat 
measurements~\cite{aoki02}. 
\begin{figure}[t]
    \begin{center}
    \includegraphics[width=70mm]{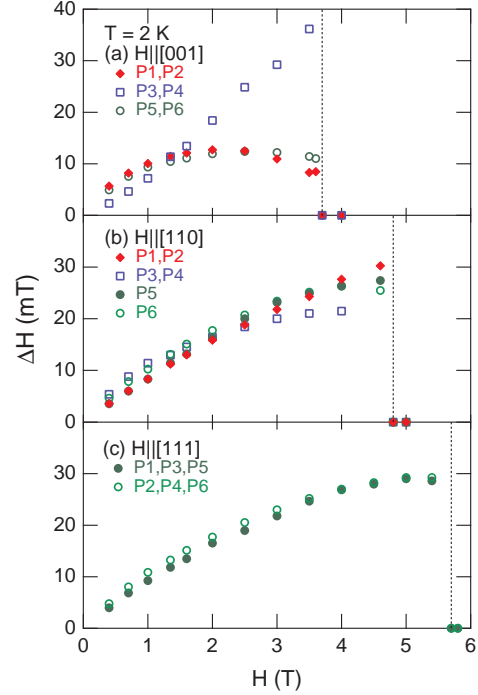}
    \caption{(Color online) Field dependence of the splitting of the hyperfine field at 2~K for the
    field along (a) [001], (b) [110], and (c) [111]. The boundary between the ordered and uniform
    phases are shown by the dotted lines.
    \label{fig:splitvsH}}
    \end{center}
\end{figure}
\begin{table}[b]
\caption{Symmetries of multipoles up to hexadecapoles in the $T_h$ crystal field. 
The + and $-$ signs show the parity under time reversal.  The multipoles are defined in terms of 
the dipole $\mathbf{J}$ as follows~\cite{shiina97, shiina04}. Quadrupoles: $O_2^0=\tfrac{1}{2}(3J_z^2-\mathbf{J}^2)$, $O_2^2=\tfrac{\sqrt{3}}{2}(J_x^2-J_y^2)$, $O_{\xi \eta}=\tfrac{\sqrt{3}}{2}\overline{J_{\xi}J_{\eta}}$, Octupoles: $T_{xyz}=\tfrac{\sqrt{15}}{6}\overline{J_xJ_yJ_z}$, $T_\xi^\alpha=\tfrac{1}{2}(2J_\xi^3-\overline{J_{\xi}J_\eta^2}-\overline{J_\zeta^2J_\xi})$, $T_\xi^\beta=\tfrac{\sqrt{15}}{6}(\overline{J_\xi J_\eta^2}-\overline{J_\zeta^2 J_\xi})$, Hexadecapoles: $H^0=\sqrt{\tfrac{175}{48}}(J_x^4+J_y^4+J_z^4-\tfrac{3}{5}\mathbf{J}^2(\mathbf{J}^2-\tfrac{1}{3}))$, $H_{\xi}^{\alpha}=\tfrac{\sqrt{35}}{8}(\overline{J_\eta^3 J_\zeta}-\overline{J_\eta J_\zeta^3})$, $H_\xi^\beta=\tfrac{\sqrt{5}}{8}(2\overline{J_\xi J_\eta^2 J_\zeta}-\overline{J_\eta^3 J_\zeta}-\overline{J_\eta J_\zeta^3})$. 
Here $\xi$, $\eta$, $\zeta$ represent $x$, $y$, $z$ and their cyclic permutation. The bars on products denote summation over all permutations of thier subscripts. Since $(T_x^\alpha,T_y^\alpha,T_z^\alpha)$ and $(T_x^\beta,T_y^\beta,T_z^\beta)$ ($(H_x^\alpha,H_y^\alpha,H_z^\alpha)$ and $(H_x^\beta,H_y^\beta,H_z^\beta)$) have the same symmetry $\Gamma_4^-$ ($\Gamma_4^+$) in the $T_h$ group, they can be mixed.}
\label{tb:multipoles1}
\begin{tabular}{ccc} \hline
    Symmetry & Magnetic multipoles & Nonmagnetic multipoles\\ \hline
    $T_{h} (\Gamma_1^+)$  & --- & $H^0$  \\
    $T_{h} (\Gamma_{23}^+)$ & --- & $O_2^0,O_2^2,H_u,H_v$ \\
    $T_{h} (\Gamma_4^+)$  & --- & $O_{xy},O_{yz},O_{zx},H_x,H_y,H_z$ \\
    $T_{h} (\Gamma_1^-)$  & $T_{xyz}$ & ---  \\
    $T_{h} (\Gamma_4^-)$  & $J_x,J_y,J_z,T_x,T_y,T_z$ & ---  \\ \hline
\end{tabular}
\end{table}
%
\begin{table*}[t]
    \caption{Symmetries of multipoles in the $T_h$ crystal field in magnetic fields. 
     The numbers in parentheses in the fifth column indicate the equivalent P sites and the last column shows the 
     number of NMR lines for a single P$_{12}$ cage in the presence of the magnetic multipoles. They should be 
     multiplied by two to get the total number of lines in the antiferro-multipole ordered phase with $\mathbf{Q}=(1,0,0)$.}
    \label{tb:multipoles2}
    \begin{tabular}{cccccc} \hline
       Field directions & Symmetry & Magnetic multipoles & Nonmagnetic multipoles & Equivalent sites & Number of lines\\ \hline
    [001] & $C_{2v} (\Gamma_1)$ & $J_z,T_z$ & $O_2^0,O_2^2,H^0,H_u,H_v$& (1,2), (3,4), (5,6) & 3\\
	   & $C_{2v} (\Gamma_2)$ &$T_{xyz}$  & $O_{xy},H_z$ & (1,2), (3,4) & 4\\ 
	   & $C_{2v} (\Gamma_3)$ & $J_x,T_x$ & $O_{zx},H_y$ & (1,2), (5,6) & 4\\ 
	   & $C_{2v} (\Gamma_4)$ & $J_y,T_y$  & $O_{yz},H_x$ & (3,4), (5,6) & 4\\ \hline
    [110] & $C_{1h} (\Gamma_1)$ & $J_x,J_y,T_x,T_y,T_{xyz}$ & $O_2^0,O_2^2,O_{xy}, H^0,H_u,H_v,H_z$  & (1,2), (3,4) & 4\\
	   & $C_{1h} (\Gamma_2)$ & $J_z,T_z$ & $O_{yz},O_{zx},H_x,H_y$  & --- & 6\\ \hline
    [111]$^\dagger$ & $C_{3} (\Gamma_1)$ & $J_c,T_c,T_{xyz}$ & $O_c,H^0,H_c$ & (1,3,5), (2,4,6) & 2\\
	   & $C_{3} (\Gamma_{23})$ & $(J_a,J_b),(T_a,T_b)$ & $(O_2^0,O_2^2),(O_a,O_b),(H_u,H_v), (H_a,H_b)$  & --- & 6 \\ \hline
    \end{tabular}
\medskip{\footnotesize{$^\dagger$$a$-, $b$- and $c$-components of the multipole 
$\Psi~(=J,O,T,H)$ are defined as $\Psi_a=\tfrac{1}{\sqrt{6}}(-\Psi_x-\Psi_y+2\Psi_z)$, 
$\Psi_b=\tfrac{1}{\sqrt{2}}(\Psi_x-\Psi_y)$, and $\Psi_c$ $=\tfrac{1}{\sqrt{3}}(\Psi_x+\Psi_y+\Psi_z)$, 
respectively, where $(\Psi_x,\Psi_y,\Psi_z)$ is either $(J_x,J_y,J_z)$, $(O_{yz},O_{zx},O_{xy})$, 
$(T_x,T_y,T_z)$ or $(H_x,H_y,H_z)$.}}
\end{table*}

We define the splitting of the hyperfine field $\Delta H$ as the frequency interval of the splitting 
divided by  the nuclear gyromagnetic ratio $\gamma_{N}$.  The field dependence of $\Delta H$ 
is plotted in Fig.~\ref{fig:splitvsH} for various field directions at $T$=2~K.  A remarkable result 
is that $\Delta H$ extrapolates to zero at $H$=0.  Such a feature has been also observed in 
polycrystals by Ishida {\it et al.}\cite{ishida04}  The field dependences of $\Delta H$ 
are different for different sites.  This is most evident for $\mathbf{H} \parallel [001]$, where $\Delta H$ for P3 and P4 increases monotonically with increasing field while $\Delta H$ for other sites 
shows a maximum.  The field variation of $\Delta H$ is smooth in all cases without any jump 
or kink, indicating absence of additional phase transitions up to the boundary to the high-field 
heavy fermion (HF) state. A first-order transition to the HF phase is apparent from the sudden 
vanishing of $\Delta H$ at the phase boundary. 

Detailed examination of angle dependence of the spectra below $T_\mathrm{A}$ reported 
in ref. \citen{sakai06} revealed that all of the NMR lines above $T_\mathrm{A}$ split always into 
two lines and $\Delta H$ never vanishes for any field direction.  Obviously, the doubling of the NMR lines 
should be ascribed to the loss of $(\tfrac{1}{2},\tfrac{1}{2},\tfrac{1}{2})$ translation and the distinct 
electronic states of the two Pr ions at the corner (PrI) and the body center (PrII) of the original bcc 
lattice~\cite{iwasa02, ishii03}.  This should divide each of the P1--P6 sites into two sublattices 
P1(I)--P6(I) and P1(II)--P6(II). The former (latter) belongs to the P$_{12}$ cage surrounding the 
PrI (PrII) sites.  We should stress that there is no additional 
line splitting, i.e., the number of NMR lines from each of these cages is exactly the same as the 
number of lines above $T_\mathrm{A}$.  This means that the $T_h$ point symmetry at both Pr sites  
is preserved below $T_\mathrm{A}$, which is not compatible with any type of quadrupole ordering.  
In the following, we provide more precise arguments on these points. 

The hyperfine field at P nuclei is determined by the spin density distribution of Pr-4$f$ electrons 
through the dipolar and the transferred hyperfine interactions. The very local nature of the latter 
interaction causes nuclei to couple not only to dipole moments but to octupoles and 
higher order magnetic multipoles.  This is because the local spin density near a P nucleus 
can be nonzero when a Pr ion has finite expectation value of a high order magnetic multipole, 
even if the dipole moment, or the spatial integration of the spin density, is zero. Such a possibility has 
been first recognized by Sakai \textit{et al.}~\cite{sakai97} in their analysis of the NMR data on   
CeB$_{6}$~\cite{takigawa83}.  

Magnetic and nonmagnetic multipoles up to hexadecapoles are presented in Table~\ref{tb:multipoles1} 
as the basis of irreducible representations of the $T_h$ group. 
In magnetic fields, they are decomposed into sets of smaller number of basis for
reduced symmetries as shown in Table~\ref{tb:multipoles2}.  The NMR line splitting $\Delta H$ 
is due to some staggered magnetic multipoles with $\mathbf{Q}=(1,0,0)$.  The vanishing 
$\Delta H$ at zero field indicates that these staggered multipoles also disappear at zero field.  
Therefore, the order parameter (OP) at zero field must be a nonmagnetic staggered multipole, which is even with respect to the time reversal and does not couple to nuclear magnetic moments.  As discussed by Shiina \textit{et al.}~\cite{shiina97}, however, OP at zero field and the field-induced multipoles must 
belong to the same irreducible representation of the symmetry group reduced by the magnetic field.  
Thus identification of the field-induced magnetic multipoles allows us to determine the symmetry of 
OP at zero field. 
\begin{figure}[b]
    \begin{center}
    \includegraphics[width=75mm]{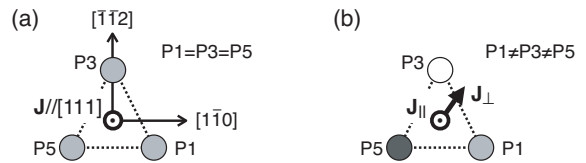}
    \caption{Field-induced dipole $\mathbf{J}$ for $\mathbf{H}\parallel [111]$. (a) When $\mathbf{J}$ is parallel to
    $\mathbf{H}$. i.e.  $J_c\neq 0$ and $(J_a, J_b)=0$ in the notation of Table~\ref{tb:multipoles2}, 
    P1, P3 and P5 are all equivalent. (b) If $J_a$ or $J_b \neq 0$, P1, P3 and P5 become inequivalent 
    due to loss of three-fold rotational symmetry about [111].
    \label{fig:perp}}
    \end{center}
\end{figure}

Using the invariant form of the hyperfine coupling at the P sites in the filled skutterrudite 
structure~\cite{sakai06,kiss06}, the difference of the hyperfine fields at P$n$(I) and P$n$(II) ($n$=1--6), 
$\Delta\mathbf{H}(n)=(\Delta H_{x}^{(n)}, \Delta H_{y}^{(n)},\Delta H_{z}^{(n)})$, is given for $n$=1 and 2 as 
\begin{align} 
            (&\pm c_{11}T_{xyz}^{s} +  c_{12}J_{x}^{s} + c_{13}T_{x}^{s},  \nonumber \\
             &c_{21}J_{y}^{s} + c_{23}T_{y}^{s} \pm (c_{22}J_{z}^{s} + c_{24}T_{z}^{s}),   \nonumber \\
            &\pm (c_{32}J_{y}^{s} + c_{33}T_{y}^{s}) + c_{32}J_{z}^{s} + c_{34}T_{z}^{s}) , 
\label{eq:Hhf}
\end{align}
where $T_{xyz}^{s}$, $T_{\xi}^{s}$, and $J_{\xi}^{s}$ are the staggered octupole and dipole
moments and $c_{ij}$ is the hyperfine coupling constants. For terms with $\pm$, the $+$ ($-$)
sign should be taken for P1 (P2). Expressions for P3 and P4 (P5 and P6) are obtained by applying once
(twice) simultaneous cyclic permutations $x \rightarrow y \rightarrow z \rightarrow x$ in the subscripts of the multipoles
and $\Delta H_{x} \rightarrow \Delta H_{y} \rightarrow \Delta H_{z} \rightarrow \Delta H_{x}$.  
Since the external field is much larger than the hyperfine field, $\Delta H$ is equal to 
$\Delta\mathbf{H} \cdot \mathbf{h}$, where $\mathbf{h}$ is the unit vector along the field direction. 
From eq.~(\ref{eq:Hhf}), we can determine the equivalent sites and the number of NMR lines in the 
presence of these field-induced magnetic multipoles as shown in the last two columns of 
Table~\ref{tb:multipoles2}.  For example, if a staggered component of $J_y$ or $T_y$ were induced by
the field along [001], eq.~(\ref{eq:Hhf}) tells $\Delta H_{z}^{(1)} \ne \Delta H_{z}^{(2)}$.  Thus 
P1 and P2 should give distinct NMR lines and there should be four lines from a single cage.     

As mentioned before, the number of NMR lines for a single P$_{12}$ cage remains unchanged 
upon entering into the ordered phase: three, four and two for $\mathbf{H}\parallel [001]$, [110] and [111], 
respectively.  From Table~\ref{tb:multipoles2}, we can conclude that the field-induced magnetic multipoles 
must have the $\Gamma_1$ symmetry for all field directions. In other words, 
only the totally-symmetric multipoles can avoid additional line splitting.  For dipoles, this means 
that the induced dipole moment is always parallel to the external field, which is consistent with the neutron 
scattering measurements.\cite{hao03}  This can be intuitively understood as illustrated in 
Fig.~\ref{fig:perp} for the case $\mathbf{H}\parallel [111]$.  

Thus the OP must be totally symmetric in magnetic fields. Inspecting Table~\ref{tb:multipoles2}, 
we find there is no such quadrupole.  For example, $\Gamma_4$-type quadrupoles 
$O_{xy}$, $O_{yz}$ and $O_{zx}$ induce magnetic multipoles which do not belong to 
$\Gamma_1$ for $\mathbf{H}\parallel [001]$. Quadrupoles $O_2^0$ and $O_2^2$ of 
$\Gamma_{23}$ type induce dipole moments perpendicular to the field for 
$\mathbf{H}\parallel [111]$.  Both cases lead to more NMR lines than experimentally observed. 
Therefore, there should not be antiferro-quadrupole order at zero field.  
\begin{table}[t]
    \caption{Reduction of irreducible representations of the $T_h$ group in magnetic fields.}
    \label{tb:reduction}
    \begin{tabular}{cccc} \hline
	zero field & $\mathbf{H}\parallel [001]$ & $\mathbf{H}\parallel [110]$ & $\mathbf{H}\parallel [111]$ \\ 
	$T_h$ & $C_{2v}$ & $C_{1h}$ & $C_3$ \\ \hline
	$\Gamma_1^\pm$ & $\Gamma_1$ & $\Gamma_1$ & $\Gamma_1$ \\
	$\Gamma_{23}^\pm$ & $2\Gamma_1$ & $2\Gamma_1$ & $\Gamma_{23}$ \\
	$\Gamma_4^\pm$ & $\Gamma_2+\Gamma_3+\Gamma_4$ & $\Gamma_1+2\Gamma_2$ &
	$\Gamma_1+\Gamma_{23}$ \\ \hline
    \end{tabular}
\end{table}

One notices at this point that the OP at zero field should be also totally symmetric.  
We can see this formally by studying reduction of irreducible representations of the $T_h$ 
group under the [001], [110] and [111] fields as shown in Table~\ref{tb:reduction}.
It is apparent that only the $\Gamma_1$ representation at zero field has always a $\Gamma_1$ component
in magnetic fields irrespective of the field direction. 

At present, we cannot identify the detailed form of the OP because the number of NMR lines depends 
only on the symmetry of the OP in magnetic fields. The OP may thus include various multipoles of 
different ranks in general, as far as they are nonmagnetic and have the $\Gamma_1$ symmetry. 
Such a multipole order can be caused, for example, by alternate breathing of the icosahedral cages which 
preserves local $T_h$ symmetry at the Pr sites. This type of lattice distortion 
has been observed in the insulating phase of PrRu$_{4}$P$_{12}$,\cite{lee04} where  
the antiferro order of the hexadecapole $H^0\propto J_x^4+J_y^4+J_z^4$ has been proposed~\cite{takimoto06}.  
In fact, this is the only totally-symmetric nonmagnetic multipole among those listed in Table~\ref{tb:multipoles1}.  
Quantitative analyses of the splitting $\Delta H$ as a function of direction and magnitude of the field
will give further information about the OP such as the rank of a dominant multipole. 
Some phenomenological approaches are in progress,~\cite{kiss06,sakai06} although a 
microscopic theory is needed to uncover the mechanism of this interesting phase transition.

In conclusion, we have presented $^{31}$P NMR data in a single crystal of PrFe$_{4}$P$_{12}$, 
confirming antiferro order of nonmagnetic multipoles at low fields. There exist field-induced staggered 
magnetic multipoles, which is compatible only with the totally-symmetric order parameter at zero field. 
The $T_h$ point symmetry at the Pr sites is thus preserved in the ordered phase, excluding any type of quadrupole order.

We thank O. Sakai, R. Shiina, A. Kiss, Y. Kuramoto, H. Harima, and T. Sakakibara for helpful discussions.  This work 
was supported by a Grant-in-Aid for Scientific Research in the Priority Area "Skutterudite"
(No.15072203 and No. 15072206) of the MEXT, Japan.


\begin{thebibliography}{99}
\bibitem{fn:site} Those sites which are transformed by the inversion or the $(\frac{1}{2},\frac{1}{2},
\frac{1}{2})$ translation always yield the same NMR frequency.
\end{thebibliography}
\end{document}